\newcommand{\comments}[1]{}
\documentclass[12pt,reqno]{article}
\usepackage{amsmath}
\usepackage{amsfonts}
\usepackage{amssymb}
\usepackage{graphics}
\usepackage{graphicx-psmin}
\usepackage{fancyhdr}
\usepackage{textpos}

\usepackage[margin=2.7cm]{geometry}
\fancypagestyle{plain}{%
\fancyhead{} 
}
\author{Edward B. Baker III\\ \textit{Department of Particle Physics and Astrophysics} \\ \textit{The Weizmann Institute of Science, Rehovot 76100, Israel}}
\title{Supersymmetric Open Wilson Lines}

\begin{document}
\maketitle
\begin{textblock*}{5cm}(11.3cm,-10.8cm)
  WIS/02/11-FEB-DPPA
\end{textblock*}
\paragraph*{Abstract}
\ In this paper we study Open Wilson Lines (OWL's) in the context of two Supersymmetric Yang Mills theories.  First we consider four dimensional N=2 Supersymmetric Yang Mills Theory with hypermultiplets transforming in the fundamental representation of the gauge group, and find supersymmetric OWL's only in the superconformal versions of these theories.  We then consider four dimensional N=4 SYM coupled to a three dimensional defect hypermultiplet.  Here there is a semi-circular supersymmetric OWL, which is related to the ray by a conformal transformation.  We perform a perturbative calculation of the operators in both theories, and discuss using localization to compute them non-perturbatively.  
\newpage
\section{Introduction}
The study and calculation of supersymmetric Wilson loops has been an interesting and fruitful area of current research.  Supersymmetric Wilson loops were originally defined in N=4 Super Yang Mills Theory (SYM) as a supersymmetric generalization of the usual Wilson loop operator, defined in Euclidean space by the following equation
\begin{equation}
\text{W(C)}=\text{Tr}(P\exp[\oint_C (dx^\mu iA_\mu(x_\nu) + |dx|\Phi_I(x_\nu) \Theta^I(x_\nu))]),
\end{equation}
where here the $\Phi_I$ are the six scalars in N=4 SYM and $\Theta_I$ is a six dimensional vector of unit length.  These loops are interesting because they maintain some of the important properties of the Wilson loop operator, such as the relationship to confinement, while their supersymmetry allows for exact calculation in some cases. 
\\ \indent In \cite{Zarembo}, it was conjectured that the circular supersymmetric Wilson loop conformally related to the straight line could be calculated at strong coupling.  It was known previously that the straight supersymmetric Wilson line has vanishing perturbative corrections.  In the case of the circular loop, it was shown that the only diagrams contributing at first order in perturbation theory are ``rainbow" and ``ladder'' diagrams, which have no interactions on the interior of the loop.  Considering only these diagrams, it was shown that the behavior of the operator at strong coupling is exactly as predicted by the ADS/CFT correspondence \cite{Maldacena:1998im}, \cite{Rey:1998ik}.  Furthermore, this behavior is identical to a prediction of a specific Hermitean Matrix Model.   They conjectured that the cancellations found at one loop order would hold at all orders, and that the value they found with the rainbow and ladder diagrams would be exact.  
\\ \indent   In \cite{DrukkerGross} it was shown that the conjecture of \cite{Zarembo} was correct, and held to all orders in the 1/N expansion.  Using the large conformal transformation relating the circle to the line, they found that there is a conformal anomaly which changes the expectation value of the operator.  This anomaly comes from the point which is taken to infinity, and therefore the calculation can be reduced to a zero dimensional QFT, or matrix model, as conjectured in \cite{Zarembo}.  They did not prove that the matrix model was Gaussian, however.
\\ \indent In \cite{Pestun}, the technique of localization was used to calculate the operator exactly, proving the conjecture, and extending the calculation to cases which had not been considered previously.  In particular, the calculation was also valid for the N=2 theory with a massive adjoint hypermultiplet, and could be done for multiple Wilson lines in the same correlator.  For these more general cases the effects of instantons were also considered, which had not been previously possible.  All of these calculations were done on a four sphere.  In \cite{Pestun:2009nn}, a new class of one eighth BPS Wilson loops (introduced in \cite{Drukker:2007dw}) were localized to a two-dimensional Hitchin-Higgs model. 
\\ \indent In this paper, we try to extend these calculations to Open Wilson Line operators.  These operators are defined by the equation
\begin{equation} \label{eq:OWLdef}
OW_i ^j [\tilde{C}]=\bar{\psi}^j(x_j)\textit{P}\exp \Big[ \int_{\tilde{C}} ds \,\big (iA_\mu (x^\nu (s))\dot{x}^\mu (s) + n^k (s) \Phi_k (x^\nu(s))|\dot{x}|(s)\big) \Big]\psi_i(x_i),
\end{equation}  
where the fermions (or scalars) at the ends of the line transform in the fundamental and anti-fundamental representations of the gauge group.  This is a fairly general class of operators, and they can be constructed in all of the theories discussed above when appropriate matter is introduced.  In \cite{Aharony:2008an}, the holographic duals of these operators were found in the context of the Sakai-Sugimoto model of QCD, and some other theories.  In the Sakai-Sugimoto model, the Open Wilson Line is related to the chiral condensate that breaks the chiral symmetry of the theory, and so it is interesting to study in that case.  Since the Sakai-Sugimoto model is not supersymmetric, the technique of localization cannot be used to compute the operator there, so we attempt to compute it in other models which are supersymmetric in order to shed some light on its behavior.  
\\ \indent To start with, we consider four-dimensional N=2 supersymmetric Yang Mills theory with a hypermultiplet transforming in the fundamental representation of the gauge group.  The calculation of the supersymmetry transformations of these operators is found in section two.  It is shown that there are no non-trivial linear supersymmetric OWL's in pure supersymmetric theories.  In superconformal theories, there are linear combinations of supersymmetry and superconformal generators which annihilate the line.  The divergences in these operators cancel in a similar fashion to the Wilson loop of N=4 SYM, and therefore may be a good candidate for localization in the future.
\\ \indent In section three, we consider another scenario originally posed in \cite{Aharony:2008an}.  This scenario is built from the theory defined in \cite{Defect}.  This is a defect conformal field theory with a three-dimensional fundamental hypermultiplet coupled to four-dimensional N=4 SYM.  We first consider a Wilson line stretched between two defects, which is found to be supersymmetric, but has a vanishing expectation value.  We then consider the semi-circular Wilson line related to the semi-infinite ray by an inversion, and find that it has a non-trivial expectation value.  This operator was computed at strong coupling in \cite{Aharony:2008an}; here we calculate this operator to first non-trivial order perturbatively.  It is found to have a logarithmic divergence, but this divergence can be interpreted simply as a wavefunction renormalization.    
\\ \indent It is likely that neither of the operators found will be localized to a vector-matrix model, because it seems that there is not enough Bosonic symmetry in the theory to properly localize it to a point.  A more likely candidate is that it will localize to a one or two dimensional theory, as in \cite{Pestun:2009nn}.  We briefly discuss this point at the end of section 3.  We end in section four with conclusions.  Three appendices contain technical details.

\section{N=2 SYM with hypermultiplet}
In this section we will consider 4D N=2 SYM with a fundamental hypermultiplet.  We will be using the conventions of \cite{Sohnius}.  For this theory, the gauge multiplet consists of a gauge field $A_\mu$, an SU(2) doublet of Majorana fermions $\lambda_i$, two scalar singlets M and N (which can be combined into a complex scalar), and an auxiliary triplet of scalars $D_m$.  The hypermultiplet consists of a doublet of two complex scalars B$_i$ and a singlet Dirac fermion $\Psi$.  The Lagrangian is then
\begin{eqnarray}
\frac{1}{g^2}\begin{LARGE}
(
\end{LARGE}\frac{1}{2}\nabla_\mu B^\dagger_i \nabla^\mu B^i + i \bar{\Psi}\gamma^\mu\nabla_\mu \Psi +iB^{\dagger i}\bar{\lambda}_i \Psi - i \bar{\Psi} \lambda^i B_i \nonumber \\ - \bar{\Psi}(M-\gamma_5 N)\Psi - \frac{1}{2}B^{\dagger i}(M^2+N^2)B_i + \frac{1}{2}B^{\dagger i}\tau_{ij}^m D_m B^j\begin{LARGE}
)
\end{LARGE}, 
\end{eqnarray}
where we have not included the Langrangian for the gauge multiplet since it will not be relevant.

\subsection{Fermion Wilson Line}
We will start by considering a Wilson line with fundamental fermions from the hypermultiplet at either end, given by the expression 
\begin{equation} \label{eq:OWLdef2}
OW_i ^j [\tilde{C}]=\bar{\Psi}^j(x_1)\textit{P}\exp \Big[ \int_{\tilde{C}} ds \,\big (iA_\mu \dot{x}^\mu (s) + (v^5M+v^6N) |\dot{x}|\big) \Big]\Psi_i(x_2).
\end{equation}   
Here the contour will be a straight line between $x_1$ and $x_2$, and $v^{5,6}$ are parameters constrained to satisfy $v_5^2+v_6^2=-1$ \footnote{The negative sign is due to the conventions used}. To annihilate this OWL, we must find supercharges that separately annihilate both fermions, and the Wilson line.  At first we will consider only supersymmetry transformations, and later will consider superconformal transformations.  In terms of Weyl spinors, the supersymmetry transformations of the fermions are\footnote{These were derived from the transformations in \cite{Sohnius}, this derivation is shown in Appendix \ref{sec:N2Susy}}
\begin{eqnarray} 
\delta \Psi_1 = i(M-iN)(\chi B_1-\xi B_2) - i\sigma^\mu\bar{\chi} \partial_\mu  B_1 - i\sigma^\mu\bar{\xi} \partial_\mu  B_2 \\
\delta \bar{\Psi}_2 = -(M+iN)(\bar{\xi}B_1+\bar{\chi}B_2) - \bar{\sigma}^\mu \xi \partial_\mu  B_1 + \bar{\sigma}^\mu \chi \partial_\mu  B_2 .
\end{eqnarray}
We will first make some general comments about this transformation.  Let us first restrict our attention to the transformation of $\Psi_1$.  We cannot annihilate $\Psi_1$ with either of the generators $\bar{Q}_1$ or $\bar{Q}_2$, since this would result in an equation of the form
\begin{equation}
\sigma^\mu\bar{\chi}=0
\end{equation}
for all $\mu$, but since the $\sigma_\mu$ form a complete basis for hermitian 2x2 matrices, the only solution to that equation is $\bar{\chi}=0$, and similarly for the term with $\bar{\xi}$.  We also cannot form linear combinations of generators to cancel these, because the two generators transform into the different scalars $B_1$ and $B_2$.  Furthermore, we cannot form linear combinations of $\Psi_1$ and $\Psi_2$ or $\bar{\Psi}_2$, because the first parameter would violate the gauge symmetry ($\Psi_2$ transforms in the anti-fundamental of the gauge group), whereas the second would violate Lorentz invariance.  For this reason, we cannot annihilate the Fermions in \eqref{eq:OWLdef2} with any supercharges.  
\subsection{Scalar Wilson Line}
We will now consider putting fundamental scalars at the ends of the Wilson line.  We will first consider pure supersymmetry transformations, with no superconformal part.  The supersymmetry variations of the scalars in the hypermultiplet are  
   \begin{eqnarray}
\delta B_1 = 2(\chi\Psi^1 + i\bar{\xi}\bar{\Psi^2}) \\
\delta B_2 = 2(\xi\Psi^1 - i\bar{\chi}\bar{\Psi^2}). 
\end{eqnarray}
In order to annihilate a linear combination of scalars of the form $u_1B_1+u_2B_2$, the linear combination 
\begin{equation} \label{eq:STrans123}
\chi Q_1 +\xi Q_2+\bar{\chi} \bar{Q}_1 +\bar{\xi} \bar{Q}_2
\end{equation}
 must satisfy the following conditions:
\begin{equation} \label{eq:cond123}
u_1\chi=-u_2\xi, \ \ \ u_2\bar{\chi}=u_1\bar{\xi}.
\end{equation}

The first thing to notice is that the conditions on the SUSY generators do not allow for a ``Hermitian" combination in which $\chi$ and $\bar{\chi}$ are dependent, because if all of the conditions above are satisfied and the combination is Hermitian, we have
\begin{eqnarray}
u_1\chi=-u_2\xi \rightarrow u_1^*\bar{\chi}=-u_2^*\bar{\xi} \rightarrow \nonumber \\ u_1^*u_1u_2\bar{\chi}=-u_2^*u_1u_2\bar{\xi} \rightarrow |u_1|^2\bar{\xi}=-|u_2|^2\bar{\xi},
\end{eqnarray} 
and the last equation is clearly a contradiction.  This will also apply for linear combinations of SUSY and superconformal charges, since this is a local requirement.
\\ \indent  If we consider a formal linear combination of SUSY generators where $\chi$ and $\bar{\chi}$ are independent, we can find solutions, but we will not find any scalars on the other side that can form altogether a singlet of the R symmetry.  To see this, we can consider a linear combination on the other side of the form $w_1\bar{B_1}+w_2\bar{B_2}$.  The transformations in this case are 
 \begin{eqnarray}
\delta \bar{B}_1 = 2(\bar{\chi} \bar{\Psi^1} - i\xi\Psi^2) \\
\delta \bar{B}_2 = 2(\bar{\xi}\bar{\Psi^1} + i\chi \Psi^2). 
\end{eqnarray}
 The conditions for this to be annihilated are opposite to those above; $w_2\chi=w_1\xi$ and $w_1\bar{\chi}=-w_2\bar{\xi}$.  We therefore find that in order to annihilate both sides, the u's and w's must satisfy $u_1w_1=-u_2w_2$. At zero seperation, this implies that the two sides contain no elements in the singlet, because of the following calculation:
 \begin{eqnarray}
 (u_1B_1+u_2B_2)(w_1\bar{B}_1+w_2\bar{B}_2)=\nonumber\\u_1w_1B_1\bar{B}_1+u_2w_2B_2\bar{B}_2+u_1w_2B_1\bar{B}_2+u_2w_1B_2\bar{B}_1=
 \nonumber \\
 u_1w_1(B_1\bar{B}_1-B_2\bar{B}_2)+u_1w_2B_1\bar{B}_2+u_2w_1B_2\bar{B}_1. \label{eq:zeroSep}
 \end{eqnarray}
Because the symmetry is global, this calculation also applies at finite seperation, and because the operator is not a singlet of the global symmetry group it will have vanishing expectation value.
\\ \indent These considerations prove that for theories that are not superconformal, all the BPS OWL's have vanishing expectation value.  In the case of superconformal theories, there are operators which are annihilated by combinations of superconformal and SUSY charges which contain singlets.  If the gauge group is SU($N_C$), which we will assume from here on, the condition for the theory to be superconformal is to have 2$N_C$ hypermultiplets in the fundamental representation.  We will construct an example by first deriving the conditions for supercharges to annihilate the Wilson line, showing that they are consistent with the scalars, and then arguing that there is a choice of superconformal plus SUSY transformation which annihilates everything.
\\ \indent The SUSY variations of the Bosonic components of the vector multiplet look like this
\comments{\begin{eqnarray} \label{eq:gaugeTrans}
\delta A_\mu = i\bar{\zeta_i} \gamma_\mu \lambda^i \\
\delta M = i\bar{\zeta_i}  \lambda^i \\
\delta N = i\bar{\zeta_i} \gamma_5 \lambda^i
\end{eqnarray}
which yields the following for the transformations \eqref{eq:gaugeTrans}}
\begin{eqnarray}
\delta A_\mu = i(\xi \sigma_\mu \bar{\lambda_2} + \bar{\chi} \bar{\sigma}_\mu \lambda_1 + \chi \sigma_\mu \bar{\lambda_1} +\bar{\xi} \bar{\sigma}_\mu \lambda_2) \\
\delta M = \chi \lambda_2-\bar{\xi} \bar{\lambda}_1-\xi \lambda_1 +\bar{\chi} \bar{\lambda}_2\\
\delta N = i(-\chi \lambda_2-\bar{\xi} \bar{\lambda}_1+\xi \lambda_1 +\bar{\chi} \bar{\lambda}_2).
\end{eqnarray} 
We want to solve the equation (where $\delta$ refers to the generator in \eqref{eq:STrans123})
\begin{equation}
\delta(iv^\mu A_\mu + v^5 M + v^6 N )= 0,
\end{equation}
which gives, after using \eqref{eq:WeylCon}, and setting the coefficients of each spinor to zero separately
\begin{eqnarray}
v^\mu \xi \sigma_\mu = (v_5+iv_6)\bar{\chi}, \ \ \ v^\mu \chi \sigma_\mu = -(v_5+iv_6)\bar{\xi} \label{eq:TransT1} \\
v^\mu \bar{\chi} \bar{\sigma}_\mu  = (-v_5+iv_6)\xi, \ \ \ v^\mu \bar{\xi} \bar{\sigma}_\mu  = (v_5-iv_6)\chi. \label{eq:TransT2}
\end{eqnarray}
It can be checked that these conditions are consistent with the SUSY conditions on the scalars \eqref{eq:cond123}, and relate different spinors so can be satisfied simultaneously.  
\\ \indent We can now argue that there must exist a linear combination of supercharges and superconformal charges which can annihilate a Wilson line connecting two fundamental scalars.  These spinors have the form $\epsilon = \epsilon_s + x^\mu \gamma_\mu \epsilon_c$, where $\epsilon_s$ is the SUSY charge and $\epsilon_c$ is the superconformal charge\footnote{Here we have absorbed any global symmetry transformation on $\epsilon_c$ into the definition of $\epsilon_c$}.  In this notation, the SUSY transformations remain the same, but are now x-dependent.  As proved above, we can choose a constant spinor $\epsilon_1$ on one side and a spinor $\epsilon_2$ on the other side of the line such that both of them annihilate their respective scalars $u_1B_1+u_2B_2$ and $w_1\bar{B}_1+w_2\bar{B}_2$ as well as the Wilson line at those points.  If we choose the first side to be at x=0 and the second at a specified point, without loss of generality we can choose (0,0,0,L), then the spinor 
\begin{equation}
\epsilon(x) \equiv \epsilon_1 +\gamma_3  x_3 \gamma_3 \frac{(\epsilon_2 - \epsilon_1)}{L}
\end{equation}
 agrees on both sides with what we want.  Furthermore, this spinor annihilates the line, because both $\epsilon_1$ and $\epsilon_2$ independently annihilate the line.  To see how this BPS condition manifests itself in perturbation theory, we next consider the pertubative corrections to the operator
 \begin{equation}\label{eq:N2OWLDef}
\frac{1}{N_C} B_1 P \exp[\int_C (iA_\mu \dot{x}^\mu + M |\dot{x}|)]\bar{B}_1,
 \end{equation} 
 where here we will be considering the straight line, and we have chosen particular scalars for convenience.  The factor of $N_C$ is used to normalize the gauge field indices. We will show that the divergences of this operator cancel as in \cite{Zarembo}, leaving a finite piece.  It is also possible to do a conformal transformation on the line, bringing it to an arc of a semi-circle, with no singular point.  This means that our analysis will also apply to the semi-circular Wilson line.  
\subsection{Cancellation of Divergences}
\begin{figure} 
\begin{center}
\includegraphics[scale=.9]{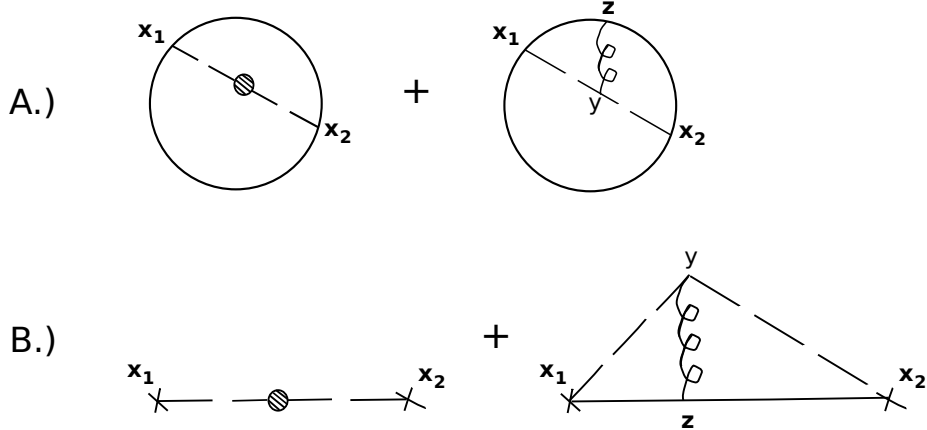} 
\end{center}
\caption{A.) Two of the diagrams that are shown to cancel in \cite{Zarembo} \indent B.) The analagous diagrams for the straight Wilson line}
\label{fig:CancelDiag}
\end{figure}

We have seen that in superconformal theories, an OWL with any choice of scalars at the ends will be annihilated by some linear combination of supercharges and superconformal charges.   We will try to see how this manifests itself in perturbation theory to order $g^4$.  At first order, the contribution is from the free propagator between the two scalars, $\Delta_{B\bar{B}}(x)=\frac{g^2N_C}{4\pi^2N_Cx^2}$.  At order $g^4$, there are two contributions to the free propagator, and the Wilson line itself receives no corrections.  The contributions are depicted in figure \ref{fig:CancelDiag}B, the first being self-energy corrections and the second being a gauge interaction between the Wilson line and the scalar propagator. 
\\ \indent  The diagrams contributing at order $g^4$ are analagous to those analyzed in \cite{Zarembo} for closed Wilson Loops in the N=4 SYM theory and depicted in figure \ref{fig:CancelDiag}A, where their divergences are found to cancel.  We will see that the divergences in \ref{fig:CancelDiag}B cancel analagously.  This can be argued in the following way:
\\ \indent  The interactions of the fundamental hypermultiplet are the same as those of an adjoint hyper-multiplet, simply changing the representation. In the computations of \cite{Zarembo}, we can view the scalars of N=4 SYM as coming from an adjoint hypermultiplet coupled to an adjoint vector multiplet.  This implies that the self-energy corrections to the scalar propagator are the same as those found in \cite{Zarembo}, up to group theory factors.  The contribution from the self-energy is therefore given in dimensional regularization by
\begin{equation}
-\delta^{ab}g^4C_2(r)\frac{\Gamma^2(\omega-1)}{2^5\pi^{2\omega}(2-\omega)(2\omega-3)}\frac{1}{[x^2]^{2\omega-3}},
\end{equation}
where $\omega\equiv \frac{d}{2}$ and $C_2(r)$ is the quadratic Casimir operator of the fundamental representation, denoted by r.  We get this factor from the term $\frac{1}{N_C}Tr(T^aT^a)$.  In the case of SU(N), $C_2(r)=\frac{N^2-1}{2N}$.  
\\ \indent The second diagram is given in Feynman gauge by the expression 
\begin{equation} \label{eq:diag2N2}
-g^4 C_2(r) \int_C d\tau \frac{dz(\tau)}{d\tau} \cdot (\frac{\partial}{\partial x^{(1)}}-\frac{\partial}{\partial x^{(2)}})\int d^{2\omega}y \Delta(x^{(1)}-y)\Delta(z-y)\Delta(x^{(2)}-y),
\end{equation}
which is exactly analagous to the expression found in \cite{Zarembo}, except for the integration contour for z and the fact that $x^{(1)}$ and $x^{(2)}$ are fixed.  It is shown there that this diagram cancels pointwise with the self-energy corrections.  The value of the divergent quantity is the same as that found in \cite{Zarembo}, although this arises from a cancellation of two factors of two.  There is a factor of $\frac{1}{2}$ since the divergence comes from the limit $z\rightarrow x^{(1)}$, and in the case of the OWL the contour ends at the points $x^{(1)}$ and $x^{(2)}$, which therefore only contributes one side of the short-distance divergence of the expression when integrated over the full circle\footnote{It can also be seen explicitly that this result is independent of the shape of the contour in between $x^{(1)}$ and $x^{(2)}$, because the divergence is local and independent of the angle of the line}.  The second factor of two comes from gauge theory contributions.  In \eqref{eq:diag2N2} the contributions from the two partial derivatives add together to yield a factor of two.  In \cite{Zarembo}, these contributions cancel with a factor of two from the gauge group structure, arising from the following calculation\footnote{We will not go into the details of the derivation, they can be reproduced from the calculations in \cite{Zarembo}}
\begin{equation}
f_{abc}\text{Tr}(T^aT^bT^c)=\frac{1}{2}f_{abc}\text{Tr}([T^a,T^b]T^c)=\frac{i}{2}f_{abc}f^{abd}\text{Tr}(T_dT^c)=\frac{i}{2}C_2(A)C_2(r)N_c.
\end{equation}
The factor of one-half in front is not present for the self-energy diagram contributions, and therefore cancels the extra one-half from the end of the Wilson line.  This means the divergences also cancel here.  It would be interesting to calculate the finite contribution to this operator, but in the next section we will do a similar computation for an operator whose expectation value is known at strong coupling.
\comments{
\subsubsection{Finite Contribution}
To calculate the finite contribution for a Wilson line which forms part of an arc of a circle, we can do an analysis following directly from \cite{Zarembo}, which we will follow closely in this section.  To calculate the value directly from a straight line segment would be a non-trivial check of this analysis.  We can, however, find the value that the straight line should have based on the finite conformal transformation from the semi-circle.  To calculate the order $g^4$ contribution from the semi-circle, we first rewrite \eqref{eq:diag2N2} using Feynman parameters and integrating over y as:
\begin{eqnarray}
g^4C_2(r)\frac{\Gamma(2\omega-2)}{2^4\pi^{2\omega}}\int_0^1 d\alpha \, d\beta \, d\gamma \ \delta(\alpha+\beta+\gamma-1)(\alpha\beta\gamma)^{2-\omega}\alpha  \\ \int_C d\tau \nonumber \frac{((1-\alpha) x_1 +\beta z + \gamma x_2)\cdot\dot{z}}{(\alpha \beta (x_1- z)^2 + \alpha \gamma (x_1- x_2)^2 + \beta \gamma (z -x_2)^2)^{2\omega-2}},
\end{eqnarray}
in the case of the a circular arc of radius R paramaterized by angle $\tau$, it proves convenient to introduce the variables $\tau_{ij}=\tau_i-\tau_j$, which satisfy the equations $|x^{(i)}|^2=R^2$,$|x^{(i)}-x^{(j)}|^2=2R^2(1-\cos(\tau_{ij}))$, $\dot{x}^{(i)}\cdot x^{(j)}=R^2\sin(\tau_{ij})$, and $\dot{x}^{(i)}\cdot \dot{x}^{(j)}=R^2\cos(\tau_{ij})$, yielding
\begin{eqnarray} \label{eq:poontang}
g^4C_2(r)R^{-2}\frac{\Gamma(2\omega-2)}{2^4\pi^{2\omega}}\int_0^1 d\alpha \, d\beta \, d\gamma \ \delta(\alpha+\beta+\gamma-1)(\alpha\beta\gamma)^{2-\omega}  \\ \int_C d\tau \nonumber \frac{(\alpha(1-\alpha)\sin(\tau_{12}) + \alpha\gamma \sin(\tau_{23}))}{(\alpha \beta (1-\cos(\tau_{12})) + \alpha \gamma (1-\cos(\tau_{23})) + \beta \gamma (1-\cos(\tau_{13})))^{2\omega-2}},
\end{eqnarray}
we will denote the denominator in \eqref{eq:poontang} as $\Delta$.  Using the identity
\begin{equation}
\int d\tau_2 \frac{1-\cos(\tau_{13})}{}
\end{equation}

We can rewrite \eqref{eq:diag2N2} using Feynman parameters and integrating over y as:

Must also include factors from propagators
\begin{eqnarray}
-g^4 C_2(r) \int_C d\tau \frac{dz(\tau)}{d\tau} \cdot (\frac{\partial}{\partial x^{(1)}}-\frac{\partial}{\partial x^{(2)}})\int d^{2\omega}w \Delta(x^{(1)}-w)\Delta(z-w)\Delta(x^{(2)}-w)=\\
2(\omega-1) g^4 C_2(r) \int_C d\tau \frac{dz(\tau)}{d\tau} \cdot \int d^{2\omega}y\frac{(x-y)_\nu}{(x^1-y)^{2\omega}}\frac{1}{(z-y)^{2\omega-2}}\frac{1}{(x^2-y)^{2\omega-2}}=\\
\frac{\Gamma(3\omega-2)}{\Gamma(\omega)\Gamma^2(\omega-1)}2(\omega-1) g^4 C_2(r) \int_C d\tau \frac{dz(\tau)}{d\tau} \cdot \int_0^1 d\alpha \, d\beta \, d\gamma  \delta(\alpha+\beta+\gamma-1)  \int d^{2\omega}y \frac{(\alpha\beta\gamma)^{\omega}\alpha(x-y)_\nu}{(\alpha(x^1-y)^2+\beta(z-y)^2+\gamma(x^2-y)^2)^{3\omega-2}}=\\
\frac{\Gamma(3\omega-2)}{\Gamma(\omega)\Gamma^2(\omega-1)}2(\omega-1) g^4 C_2(r) \int_C d\tau \frac{dz(\tau)}{d\tau} \cdot \int_0^1 d\alpha \, d\beta \, d\gamma  \delta(\alpha+\beta+\gamma-1)  \int d^{2\omega}y \frac{(\alpha\beta\gamma)^{\omega}\alpha(x-y)_\nu}{(y'^2+(\alpha \beta (x_1- z)^2 + \alpha \gamma (x_1- x_2)^2 + \beta \gamma (z -x_2)^2))^{3\omega-2}}=\\
\frac{\Gamma(3\omega-2)}{\Gamma(\omega)\Gamma^2(\omega-1)}2(\omega-1) g^4 C_2(r)\int_0^1 d\alpha \, d\beta \, d\gamma  \delta(\alpha+\beta+\gamma-1) \frac{(\alpha\beta\gamma)^{2-\omega}\alpha((1-\alpha) x_1 +\beta z + \gamma x_2)_\nu}{(\alpha \beta (x_1- z)^2 + \alpha \gamma (x_1- x_2)^2 + \beta \gamma (z -x_2)^2)^{2\omega-2}}
\end{eqnarray}
\begin{eqnarray}
(\alpha(x^1-y)^2+\beta(z-y)^2+\gamma(x^2-y)^2)^4=\\
y^2-2y(\alpha x_1 +\beta z + \gamma x_2) + (\alpha x_1^2 + \beta y^2 + \gamma x_2^2) = \\
(y-(\alpha x_1 +\beta z + \gamma x_2))^2-(\alpha x_1 +\beta z + \gamma x_2)^2+ (\alpha x_1^2 + \beta y^2 + \gamma x_2^2)=\\
(y-(\alpha x_1 +\beta z + \gamma x_2))^2+(\alpha \beta (x_1- z)^2 + \alpha \gamma (x_1- x_2)^2 + \beta \gamma (z -x_2)^2)
\end{eqnarray}
\begin{eqnarray}
g^4C_2(r)\frac{\Gamma(2\omega-2)}{2^4\pi^{2\omega}}\int_0^1 d\alpha \, d\beta \, d\gamma \ \delta(\alpha+\beta+\gamma-1)(\alpha\beta\gamma)^{2-\omega}\alpha  \\ \int_C d\tau \nonumber \frac{((1-\alpha) x_1 +\beta z + \gamma x_2)\cdot\dot{z}}{(\alpha \beta (x_1- z)^2 + \alpha \gamma (x_1- x_2)^2 + \beta \gamma (z -x_2)^2)^{2\omega-2}},
\end{eqnarray}
where here z is paramaterized implicitly by $\tau$.  Here we will change basis so that $x_1=0$ and $x_2=L$ and the contour is a line from $x_1$ to $x_2$ parameterized by $z=L\tau$. We get
\begin{eqnarray} \label{eq:nuts}
g^4C_2(r)L^{6-4\omega}\frac{\Gamma(2\omega-2)}{2^4\pi^{2\omega}}\int_0^1 d\alpha \, d\beta \, d\gamma \ \delta(\alpha+\beta+\gamma-1)(\alpha\beta\gamma)^{2-\omega} \\ \int_C d\tau \nonumber  \frac{(\beta \tau + \gamma )\alpha}{(\alpha \beta \tau^2 + \alpha \gamma + \beta \gamma (\tau -1)^2)^{2\omega-2}},
\end{eqnarray}
here we will use a trick similar to the one found in \cite{Zarembo}.  Denoting the denominator of \eqref{eq:nuts} as $\Delta$, we consider the identity
\begin{eqnarray}
\int_0^1 d\tau \frac{\partial}{\partial \tau} \frac{(1-\tau)}{\Delta^{2\omega-3}}=0,
\end{eqnarray}
which can be expanded as 
\begin{eqnarray}
-\int_0^1 d\tau \left(\frac{1}{\Delta^{2\omega-3}}+(2\omega-3)\frac{2\beta(1-\tau)((1-\beta)\tau-\gamma )}{\Delta^{2\omega-2}}\right)=\frac{1}{(\gamma(1-\gamma))^{2\omega-2}}.
\end{eqnarray}
Unlike in \cite{Zarembo}, we will not have full cancellation with the self-energy diagrams, but we do expect to get something finite.  Since we have isolated the divergence on the RHS, which should cancel with the self-energy diagrams (based on the previous arguments), we can now restrict the rest of the identity to $\omega=2$.  We rewrite the identity as
\begin{eqnarray}
\int_0^1 d\tau\frac{\alpha(\beta \tau + \gamma)}{\Delta^2}=\int_0^1 d\tau \frac{-\tau^2\beta(1-\beta)+(\alpha\beta+2\beta(1-\beta))\tau +(2\alpha\gamma-\beta\gamma))}{\Delta^2}\nonumber \\+\frac{1}{(\gamma(1-\gamma))^{2\omega-2}}.
\end{eqnarray}
We will now try to evaluate the first term on the right hand side.  To do this, we can complete the square in the denominator
\begin{eqnarray}
\alpha \beta \tau^2 + \alpha \gamma + \beta \gamma (\tau -1)^2 = \\
\beta(1-\beta)(\tau-\frac{\gamma}{1-\beta})^2-\beta \frac{\gamma^2}{1-\beta}+\gamma(1-\gamma)= \\
\beta(1-\beta)(\tau-\frac{\gamma}{1-\beta})^2+\gamma\frac{1-\alpha}{1-\beta}
\end{eqnarray}

\begin{eqnarray}
g^4C_2(r)L^{6-4\omega}\frac{\Gamma(2\omega-2)}{2^4\pi^{2\omega}}\int_0^1 d\alpha \, d\beta \, d\gamma \ \delta(\alpha+\beta+\gamma-1)(\alpha\beta\gamma)^{2-\omega}\alpha  \\ \int_{-\chi L}^{L(1-\chi)} d\tau' \nonumber \frac{(\beta \tau'+ \chi)}{(\beta(1-\beta)\tau'^2+\chi(1-\alpha))^{2\omega-2}} 
\end{eqnarray}
where here $\chi\equiv \frac{\gamma}{1-\beta}$ and $\tau'=\tau-\chi$.  We can now perform the integral over $\tau'$ to get 

\begin{equation}
s
\end{equation}}
\comments{ 
  In this case, we can check that in order for the other side to be annihilated, we have to put the linear combination $\bar{A_1}+i\bar{A_2}$ since we have
 \begin{eqnarray}
(\chi Q_1 +i\xi Q_2)(\bar{A_1}+i\bar{A_2})=2i(-\chi+\xi) \Psi^2 \\
(\bar{\chi} \bar{Q}_1 +i\bar{\xi} \bar{Q}_2)(\bar{A_1}+i\bar{A_2})=2i(\bar{\chi}-\bar{\xi}) \bar{\Psi^1 }
 \end{eqnarray}}
\comments{This actually prohibits having a non-trivial expectation value for the OWL. All of the fields appearing in the Wilson line are singlets of the R symmetry, so the full operator contains no singlets of the R symmetry.  If the expectation value is a number, it cannot transform under the R symmetry, and must therefore vanish. 
\\ \indent Next we could consider a semi-circle instead of a line segment; now there may be superconformal symmetries which annihilate the Wilson line, and these could change the local form of the SUSY transformation (as found in the next section).  However, there is a non-singular conformal transformation which takes the finite line to the semi-circle.  Consider the line stretching from (-1,1) to (1,1).  Under an inversion around the origin, defined by the equation $x_\mu\rightarrow\frac{x_\mu}{x^2}$, it is transformed into a semi-circle, as depicted in figure \ref{fig:CT}.  Because no supersymmetries annihilate operators with SU(2) singlets for the straight line, we are guaranteed that no semi-circular operator constructed in this way can be annihilated by any linear combination of superconformal and supercharges.  In appendix \ref{sec:Divergence}, we show that this argument can be verified in perturbation theory, where there is a divergence in the operator that does not cancel as in supersymmetric operators.  The fact that this result does not depend on the precise choice of scalars suggests that there are no non-trivial supersymmetric OWL's in the theory.  
\\ \indent We have thus found that the only supersymmetric straight OWL's in this theory have trivial expectation value.  In the next section, we will consider a theory which does have a non-trivial OWL constructed in this way.  
\begin{figure} 
\begin{center}
\includegraphics[scale=1]{Conformal.eps} 
\end{center}
\caption{Conformal transformation from the line to the semi-circle.}
\label{fig:CT}
\end{figure}
}
\section{Defect Conformal Field Theory}
In this section, we will consider a different setup in order to find non-trivial SUSY Open Wilson lines.  The theory we will consider is a ``defect conformal field theory", which is dual to a D5-D3 brane configuration in string theory, and was developed in \cite{Defect}.  The field theory consists of N=4 Supersymmetric Yang Mills in four-dimensional space, and it is coupled to an N=4 supersymmetric hypermultiplet which lives in a three-dimensional flat hyperplane embedded in the space. The bulk field content is that of an N=4 multiplet, with one gauge field $A_\mu$, four Majorana spinors including the gaugino $\lambda$ and a triplet $\chi^A$, and six scalars denoted $X^V_A$ and $X^H_A$ (A$\in \{1,2,3\}$) for reasons that will become clear.  All these fields transform in the adjoint representation of the gauge group SU(N$_C$).  The defect also has N=4 supersymmetry, but the generators in three dimensions contain half the degrees of freedom of four dimensional spinors.  The defect therefore breaks the SU(4) R-symmetry of the full theory to an SO(4)$\simeq$SU(2)$_V$xSU(2)$_H$ symmetry group.  The defect matter consists of a three dimensional hypermultiplet, which is composed of an SU(2)$_H$ doublet of complex scalars $q_m$, an SU(2)$_V$ doublet of 3D Dirac fermions $\Psi^i$, and two complex auxiliary fields $f_m$.  All of these fields transform in the fundamental representation of the gauge group.  The defect action, the conventions for $\gamma$ matrices and the method of reducing spinors from 4D to 3D are described in \cite{Defect} and reviewed in Appendix \ref{sec:conventions}.

\subsection{Supersymmetry Analysis}
Here we will analyze the supersymmetry transformations of the following Wilson line
\begin{equation}
q P\exp(\int (iA_\mu dx^\mu + X_3^H ds))\bar{q},
\end{equation}
where we will specify the choices of scalars soon.  We will first consider a straight Wilson line stretched in the $x_3$ direction between two defects localized in x$_3$.  The supercharges are doublets of $SU(2)_H$, while the Wilson line includes
the singlet $A_3$ and the element $X_3^H$ of the triplet, which are both mapped by the supercharges to the fermions in the bulk hypermultiplet. 
 There exists a U(1) subgroup of SU(2)$_H$ under which X$_3^H$ is neutral, because it transforms in the triplet.  The SUSY generators, however, are doublets of SU(2)$_H$, and will therefore have either positive or
negative charge under this U(1), of magnitude $\frac{1}{2}$. Because $A_3$ and $X_3^H$ are both neutral, the positive charges transform both into the same fermion (with U(1) charge $+\frac{1}{2}$), as do the negative charges.  In Appendix \ref{sec:DCFT} it is shown that these charges correspond to the two SU(2)$_H$ indices, so that $S_{i2}$ annihilates $iA_3+X_3^H$, and $S_{i1}$ annihilates $iA_3-X_3^H$.
\\ \indent It is possible to find operators at the end of the Wilson line which are also annihilated by the same supercharges. 
Because the singlet operator $\bar{q}_m q_m$ is mapped into the operator $\bar{\Psi}_i q_m+\bar{q}_m \Psi_i$, we do not expect the singlet to be mapped to zero. In Appendix \ref{sec:DCFT} it is shown that at zero seperation, the BPS operators are members of the triplet $\bar{q}_m \sigma^I_{mn} q_n$.  Since these operators are not singlets of the R-symmetry, they must have vanishing vev.  In this case we cannot have a non-trivial expectation value anyways because the scalars on the ends live on separate defects.
\begin{figure} 
\begin{center}
\includegraphics[scale=1]{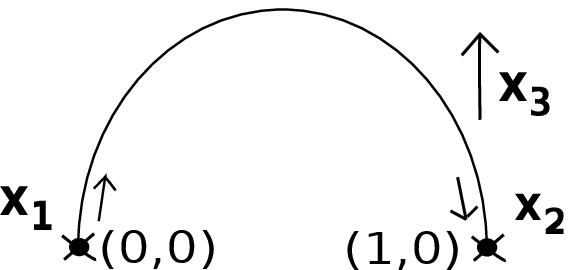} 
\end{center}
\caption{The semi-circular Wilson line}
\label{fig:sc}
\end{figure}
\\ \indent In the case of the semi-circular Wilson line depicted in figure \ref{fig:sc}, where both sides end on the same defect, the situation is different.  We first note that the semi-circle is related by a conformal transformation to a straight ray ending on the defect.  This means that the operator and supersymmetry transformations are also related by conformal and superconformal transformations to those relevant in the case of the ray (a limiting case of the case studied above).  
\\ \indent Therefore, the generators that annihilate the semi-circular line are combinations of superconformal and supersymmetry generators.  The form of the full superconformal transformations is the following, in the set of conventions used in \cite{Pestun} and \cite{Drukker:2006ga}:
\begin{eqnarray}
\delta_\epsilon A_M = \epsilon(x)\Gamma_M \lambda \\
\epsilon(x) = \epsilon_0 + \sigma_\alpha x^\alpha \epsilon_1.
\end{eqnarray} 
The form of this transformation is all that matters, so the conventions will not be important.  What is important is that locally, the superconformal transformations are the same as supersymmetry transformations.  It is only globally that they are different, since the spinor depends on x.  The same is true for the supersymmetry transformations of the fields in the defect, since this theory is also superconformal.   We can therefore say that locally, around the ends of the line, the spinors that annihilate the loop are the same as the straight Wilson line.  Globally, however, the U(1) charge changes from one end of the line to the other.  It is easy to see this because at either end of the line the sign of dx is different, so the relative sign between i$A_3$ and $X^H_3$ changes.  This means that at one end the loop is locally like a semi-infinite ray pointing away from the defect, and at the other end it is pointing towards.  Since this operator is half-BPS, the same set of supersymmetry transformations must annihilate it locally as those that annihilate the ray.
\\ \indent Because the two sides of the line are now annihilated locally by different supercharges, we see that the proper operator to consider for a semi-circular contour C as in figure \ref{fig:sc} is 
\begin{equation} \label{eq:DefOfOWL}
\text{OWL(C)}=\frac{1}{N}q_1 P \exp (\int_C(idx\cdot A + d|x| X_3^H)) \bar{q_1},
\end{equation}
since the U(1) charge of $q_1$ is opposite to that of $\bar{q}_1$, as is necessary.  The factor of $\frac{1}{N}$ is introduced to cancel the sum over the gauge index in q, where N is the dimension of the fundamental representation.  This operator can have a non-vanishing vev because it contains singlets.  We will therefore be interested in this operator since it is non-trivial. 
\comments{\\ \indent Note that unlike in the previous seciton, the only way to get the semi-circle referred to is by a conformal transformation from the infinite ray, since only conformal transformations preserving the defect are allowed.  For the semi-infinite ray, it is not clear what the operator at infinity should be, and this can only be determined through a SUSY analysis, as done above.  }
\subsection{Perturbative Calculations}
We will now evaluate \eqref{eq:DefOfOWL} perturbatively.  The first contribution to the expectation value is the free propagator between the two scalars at the end of the line, summed over the gauge index of q, which after the normalization is given by $\frac{1}{N}\frac{Ng^2}{4\pi|x_1-x_2|}$.  The free propagator will be denoted by $\Delta_{q\bar{q}}$.  There are three corrections at order $g^4$.  One comes from corrections to the Wilson line from scalar-scalar and gauge-gauge interactions, which was evaluated in \cite{DrukkerGross}.  The expression is \footnote{In \cite{DrukkerGross} the corrections are integrated to $2\pi$ since it is a full circle.}
\begin{eqnarray} \label{eq:DGresult}
\Delta_{q\bar{q}}\int_C\int_C <-A^\mu A^\nu dx^1_\mu dx^2_\nu +  X_3^H  X_3^H |dx^1||dx^2| > = \nonumber \\
 \Delta_{q\bar{q}}\int_0^{\pi} ds dt \frac{g^2C_2(r)}{8\pi^2} \frac{-\dot{x}(t)\cdot\dot{x}(s) + |\dot{x}(t)||\dot{x}(s)|}{(x(t)-x(s))^2} =\nonumber \\
  \Delta_{q\bar{q}}\int_0^{\pi} ds dt \frac{g^2C_2(r)}{8\pi^2} \frac{1}{2} =\Delta_{q\bar{q}} \frac{g^2C_2(r)}{16}, 
\end{eqnarray}
where the gauge group indices are computed in a similar way to those in the N=2 theory.  In the third line we have used the fact that for a circle $(x_1-x_2)^2=-2(\dot{x}_1\cdot\dot{x}_2 - |\dot{x}_1||\dot{x}_2|)$.  The other two corrections come from the interaction between the defect scalars and the Wilson line, as depicted in figure \ref{fig:FirstOrder}, and self-energy corrections to the defect scalar propagator.  We will consider the diagram in figure \ref{fig:FirstOrder} in the next section, and the self-energy corrections in the section after that.
\subsubsection{Gauge Interaction with Scalar Propagator}
\begin{figure} 
\begin{center}
\includegraphics[scale=1.15]{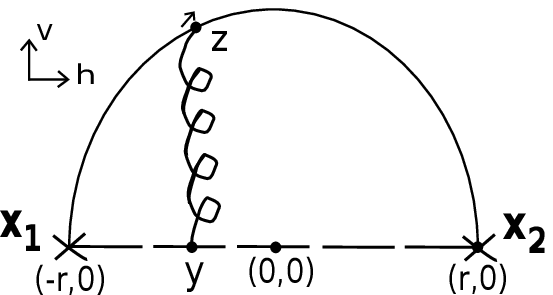} 
\end{center}
\caption{Contributing diagram to the semi-circular Wilson line at first order in perturbation theory.}
\label{fig:FirstOrder}
\end{figure}

Here we will work in position space and Feynman gauge.  The propagator for the vector field in Euclidean space is given by $\frac{g^2g_{\mu\nu}}{4\pi^2(y-z)^2}$.  For the scalar fields we get a three dimensional propagator given by $\frac{g^2}{4\pi|x-y|}$. There is a derivative which acts on either of the two 3D propagators, which gives
\begin{equation}
i\frac{\partial}{\partial x_k} \frac{1}{4\pi|x-y|} = \frac{-i(x_1-y)_k}{4\pi|x_1-y|^3}.
\end{equation}
The gauge indices again yield a factor of $\frac{1}{N}Tr(T^aT^a)=C_2(r)$, and there are no symmetry factors.  We therefore get 
\begin{equation}
\frac{2g^4C_2(r)}{(4\pi)^4}\int dz_\mu \, d^3 y \,  (\frac{(x_1-y)_k}{|x_1-y|^2}-(x_1 \leftrightarrow x_2))  \frac{1}{|x_1-y|}\frac{g^{k\mu}}{(y-z)^2}\frac{1}{|x_2-y|}.
\end{equation}
For now we will drop the constant in front, and add it at the end.  We will perform the z integral first.  We denote the vertical direction in figure \ref{fig:FirstOrder}  as v and the horizontal direction (going from $x_1$ at -r to $x_2$ at r) as h, and the other directions as $\bot$.  We then get
\begin{eqnarray}
\int d^3 y \,  (\frac{-2(x_1-y)_h}{|x_1-y|^2}-(x_1 \leftrightarrow x_2))  \frac{1}{|x_1-y|}\frac{1}{|x_2-y|}\int dz_h \,\frac{1}{y_\bot^2+(y-z)_h^2+r^2-z_h^2}, 
\end{eqnarray}
 where r is the radius of the semi-circle. The z integral gives
\begin{equation}   
\int_{-r}^{r} dz_h \,\frac{1}{y_\bot^2+(y-z)_h^2+r^2-z_h^2} =  -\frac{1}{2y_h}\text{log}(\frac{(y-x_2)^2}{(y-x_1)^2}),
\end{equation}
So the integral we must now calculate is 
\begin{eqnarray}
-\int d^3 y \,  (\frac{-(x_1-y)_h}{|x_1-y|^2}-(x_1 \leftrightarrow x_2))  \frac{1}{|x_1-y|}\frac{1}{|x_2-y|} \frac{1}{2y_h}\text{log}(\frac{(y-x_2)^2}{(y-x_1)^2}) = \nonumber \\
2\pi\int d y_\bot dy_h \,  y_\bot(\frac{(x_1-y)_h}{|x_1-y|^2}-(x_1 \leftrightarrow x_2))  \frac{1}{|x_1-y|}\frac{1}{|x_2-y|} \frac{1}{2y_h}\text{log}(\frac{(y-x_2)^2}{(y-x_1)^2}).
\end{eqnarray}
Because most of the terms depend on the distances $l_1=|x_1-y|$ and $l_2=|x_2-y|$, we will consider the change of variables $\{y_\bot,y_h\}\rightarrow\{l_1,l_2\}$.  The Jacobian is
\begin{equation*}
  \Big{|} 
  \begin{array}{cc}
  \ \partial_{\bot}l_1 & \partial_{h}l_1 \\
  \  \partial_{\bot}l_2 & \partial_{h}l_2
  \end{array} \Big{|}  ^{-1}
  =   \Big{|} 
  \begin{array}{cc}  \  \frac{y_\bot}{l_1} & \frac{y_h+r}{l_1} \\
  \  \frac{y_\bot}{l_2} & \frac{y_h-r}{l_2}
  \end{array} \Big{|}  ^{-1}  = -\frac{l_1l_2}{2y_\bot r},
 \end{equation*}
so we get (where in the first equation we keep $y_h$ for computational purposes)
\begin{eqnarray}
-\pi\int_0^\infty d l_1 \int_{|l_1-2r|}^{l_1+2r} dl_2 \, \frac{(-r-y_h)l_2^2-(r-y_h)l_1^2}{rl_1^2l_2^2y_h} \text{log}(\frac{l_2}{l_1}) \nonumber = \\
-\pi\int_0^\infty d l_1 \int_{|l_1-2r|}^{l_1+2r} dl_2 \, \frac{1}{rl_1^2l_2^2}(-4r^2\frac{l_1^2+l_2^2}{l_1^2-l_2^2}+l_1^2-l_2^2) \text{log}(\frac{l_2}{l_1}),
\end{eqnarray}
where we have used the fact that $l_1^2-l_2^2$=4r$y_h$. \\ \indent
Because this integral is symmetric with respect to ln($\frac{l_1}{l_2}$), it seems reasonable to change variables so that this is one of the variables.  The other natural choice then will be $\frac{r^2}{l_1l_2}$, where the reason for the inverse will become clear, and the r is to make everything dimensionless.  The Jacobian of the transformation is found to be $-\frac{l_1^2l_2^2}{2r^2}$, which is clearly convenient.  To find the boundary conditions, we note that in the $\{l_1,l_2\}$ plane the lines bounding the region of integration are given by 
\begin{equation}
l_1+l_2=2r, \ \ \ l_1-l_2=2r, \ \ \ l_2-l_1=2r,
\end{equation} 
but we also notice that the whole region is symmetric over the line $l_1=l_2$, and we can integrate over only one side of this region, let's say on the side where $l_1>l_2$, then multiply by two.  We define the new variables as $\theta$=ln($\frac{l_1}{l_2}$) and t=$\frac{r^2}{l_1l_2}$.  If we fix $\theta$, then the limits of integration for t are found by solving the equations
\begin{eqnarray}
\text{cosh}\left(\frac{\theta}{2}\right)=\sqrt{t} \\
\text{sinh}\left(\frac{\theta}{2}\right)=\sqrt{t},
\end{eqnarray} 
so we get 
\begin{eqnarray}
-\frac{\pi}{r} \int_{0}^{\infty} d \theta \int_{\text{sinh}^2(\theta/2)}^{\text{cosh}^2(\theta/2)} dt (-2(\text{tanh}(\theta))^{-1}+\frac{1}{t}\text{sinh}(\theta))\theta.
\end{eqnarray}
Performing the integral over t yields
\begin{eqnarray}
-\frac{\pi}{r} \int_{0}^{\infty} d \theta (-2(\text{tanh}(\theta))^{-1}+\text{ln}\left(\frac{\text{cosh}^2(\theta/2)}{\text{sinh}^2(\theta/2)}\right)\text{sinh}(\theta))\theta \nonumber = \frac{4\pi}{r}.
\end{eqnarray}
Including the overall constant, we get for this diagram
\begin{equation} \label{eq:intResult}
\frac{4g^4C_2(r)}{(4\pi)^3|x|}.
\end{equation}
We will show the overall value for all $g^4$ corrections at the end of the next section.
\subsubsection{Self Energy Corrections}
We will now evaluate the self-energy corrections to the scalar propagator, which will be the only other contributing diagrams at this order of perturbation theory.  Since the theory under consideration is super-conformal, we expect that the linear and quadratic divergence terms should cancel, and this should imply at most a logarithmic divergence in the case of the scalar self-energy.  This point is discussed in depth in \cite{Defect}.  We will work in Euclidean space.  The first self energy terms come from the vector multiplet through the gauge coupling
\begin{equation}
-iA^a_kT^aq^{m\dagger}\partial^k q^m  + h.c..
\end{equation}
In order to compute the contribution from this vertex, we note that we must use the ``pinned propagator" for the gauge fields (as discussed in \cite{Defect}), since the coupling is via the fields restricted to the defect.  The pinned vector propagator in Euclidean signature and Feynman gauge is $\frac{g^2g_{ab}}{2|k|}$, which yields (for the propagator with momentum p)
\begin{equation}
i^2g^4C_2(r)\int \frac{d^3 q}{2\pi^3} \frac{-i^2(q+2p)^2}{2|q|(p+q)^2}.
\end{equation}
We also have a contribution from the Yukawa terms
\begin{equation}
i\bar{q}^m(\bar{\lambda}_1^a)_{mi}T^a\Psi^i-i\bar{\Psi}^i(\lambda_1^a)_{im}T^aq^m.
\end{equation}
The propagators for the defect and ``pinned" fermions, respectively, are 
\begin{equation}
\frac{i\rho^k k_k}{k^2}\ , \ \ \ \ \frac{i\rho^k k_k}{2|k|},
\end{equation}
which gives
\begin{equation}
-2g^4C_2(r)\int \frac{d^3 q}{(2\pi)^3} \frac{i(p+q)_aiq_bTr(\rho^a\rho^b)}{2|q|(p+q)^2}.
\end{equation}
Here the extra two comes from the contraction over i, or in N=4 language from the four terms in the expansion \eqref{eq:lamb} for $\lambda_{im}$, which all contribute with the same sign because of the hermitian conjugation in the action, and an extra $\frac{1}{2}$ from the Majorana condition. Performing the trace yields $2g^{ab}$, giving
\begin{equation}
2g^4C_2(r)\int \frac{d^3 q}{(2\pi)^3} \frac{2(p+q)\cdot q }{2|q|(p+q)^2}.
\end{equation}
We also get a contribution from the vertex 
\begin{equation} \label{eq:Dsix}
-\bar{q}^m\sigma_{mn}^I(D_3 X_H^{Ia})T^aq^n +h.c.,
\end{equation}
which is most easily analyzed in conjunction with the four point vertices, which are
\begin{eqnarray}
A^a_kT^a q^{m\dagger}A^{bk}T^bq^m
 - \frac{1}{2}\bar{q}^m\{T^a,T^b\}q^m X_V^{Aa} X_V^{Ab} \nonumber
\\ + \frac{1}{2}\epsilon_{IJK} f^{abc} X_H^{Jb}X_H^{Kc}\bar{q}^m\sigma_{mn}^I T^a q^n  
-\frac{1}{2}\delta(0)(\bar{q}^m\sigma_{mn}^I T^a q^n)^2.
\end{eqnarray}
We first note, as in \cite{Defect}, that the terms listed above (besides the gauge coupling) come from solving the equations of motion for the auxiliary fields in the couplings 
\begin{equation}
\sigma_{ij}^A(\bar{q}^iX_V^{Aa}T^af^j+\bar{q}^iF_V^{Aa}T^aq^j +h.c.).
\end{equation}
All of the auxiliary fields have propagator -1.  We therefore get the following contributions from these terms, respectively:
\begin{equation}
-\int \frac{d^3 q}{(2\pi)^3}  \frac{3}{2|q|} -\int \frac{d^4 q}{(2\pi)^4}\frac{3}{(p+q)_{3d}^2}. 
\end{equation}
Including the terms from \eqref{eq:Dsix} and the gauge coupling yields (in Euclidean space)
\begin{equation}
g^4C_2(r)\int \frac{d^4 q}{(2\pi)^4} \frac{1}{(p+q)_{3d}^2}(-3+\frac{3q_3^2}{q^2})+\int\frac{d^3 q}{(2\pi)^3}(-\frac{3}{|q|}+\frac{3}{|q|}), 
\end{equation}
where the factor of three in the last term (the gauge coupling) comes from the trace of the metric, and the rest come from summation over the SU(2) triplet index.  To get the effective three dimensional propagator, we can integrate over the $q_3$ direction, and get the following (similar to the result found in \cite{Defect}):
\begin{equation}
g^4C_2(r)\int \frac{d^3 q}{(2\pi)^4} \frac{3q^2}{2|q|(p+q)^2}.
\end{equation}
Putting it all together, we get
\begin{equation}
g^4C_2(r)\int \frac{d^3 q}{(2\pi)^3} \frac{-4(p+q)\cdot q+(q+2p)^2+3q^2}{2|q|(p+q)^2}.
\end{equation}
As anticipated, the term quadratic in q cancels.  The remaining term is 
\begin{equation}
g^4C_2(r)\int \frac{d^3 q}{(2\pi)^3} \frac{4p^2}{2|q|(p+q)^2}.
\end{equation}
To integrate this, we switch to spherical coordinates, where $\mathbf{p}$ points in the z direction, yielding
 \begin{eqnarray}
2 g^4p^2C_2(r)\int \frac{dq d\phi}{(2\pi)^2} \frac{q \sin(\phi)}{q^2+p^2+2q|p|\cos(\phi)} \nonumber = \\
g^4|p|C_2(r)\int_0^\infty \frac{dq}{(2\pi)^2} \log \frac{(p-q)^2}{(p+q)^2} = \nonumber \\
g^4|p|C_2(r)\left(\log(q^2-p^2)+\frac{q}{|p|} \log \frac{(p-q)^2}{(p+q)^2}\right)|_{0}^{\infty}.
\end{eqnarray}
It seems that the second term above diverges linearly, but if we Taylor expand the logarithm in the limit $q\rightarrow \infty$ to first order in $\frac{1}{q}$, we get $\frac{4p}{q}$, which means that the second term is finite in the limit $q\rightarrow \infty$.  We can therefore see that the self-energy is logarithmically divergent, as expected.  As discussed in \cite{Defect}, this logarithmic divergence is related to an infinite wavefunction renormalization.  We can always choose this wavefunction renormalization so that the self-energy is renormalized to one, which is what we will do.  With this choice of renormalization, no scalar self-energy diagrams contribute to the expectation value of the OWL.
\\ \indent We conclude that after the proper renormalization, the Open Wilson line is given by 
\begin{equation}
<OWL>(g)=\frac{g^2}{4\pi|x|} (1+\frac{g^2C_2(r)}{16}(1+\frac{4}{\pi^2})+O(g^4)).
\end{equation}
\subsection{Localization Discussion}
It is unlikely that the operator \eqref{eq:DefOfOWL} will localize to a zero-dimensional vector+matrix model.  For one thing, the supersymmetry transformations close on Bosonic symmetries which preserve the defect, and therefore the operator that Pestun uses to localize in \cite{Pestun} would not work, since this rotates the Wilson loop.  Secondly, the fact that there is an infinite renormalization of the scalar field at every order of perturbation theory does not lend itself naturally to a zero dimensional vector+matrix model.  A more likely scenario is that the model will localize to a one or two-dimensional quantum field theory, as in \cite{Pestun:2009nn}.

\section{Conclusion}
 In this paper we have analyzed supersymmetric Open Wilson Lines, which are an interesting class of operators which generalize supersymmetric Wilson Loops, which have been studied and understood well in the literature.  First we considered the case of Wilson lines in N=2 SYM with a fundamental hypermultiplet, which we found to be supersymmetric only in the case of superconformal theories, and we showed that the divergences at order $g^4$ cancel in a similar fashion to those found in \cite{Zarembo}.  We then extended the analysis to the theory described in \cite{Defect}, which is dual to a string theory scenario with a D3-D5 brane intersection.  In this scenario, there are non-trivial supersymmetric semi-circular Wilson lines with scalars at the end.  We found the perturbative expansion of this operator to first non-trivial order.   It would also be interesting to extend these calculations to fermions.  Furthermore, it would be useful to see if this operator can be localized using the techniques of \cite{Pestun}, or more likely those of \cite{Pestun:2009nn}.  There may even be a supersymmetry generator which falls into the subclass of those localized in \cite{Pestun:2009nn}, which would then reduce the problem to localizing the defect theory. The expectation value of this operator at strong coupling was found in \cite{Aharony:2008an} on the string theory side of the duality proposed in \cite{Defect}, and so any expression derived can be compared to this value. These possibilities require further investigation.
 \begin{center}
\textbf{Acknowledgments}
\end{center} 
I would like to thank Ofer Aharony for many interesting discussions and for collaboration, and to thank Nadav Drukker for useful discussions. This work was supported in part by the Israel$-$U.S. Binational Science Foundation, by a research center supported by the Israel Science
Foundation (grant number 1468/06), by a grant (DIP H52) of the German Israel
Project Cooperation, and by the Minerva foundation with funding from the Federal
German Ministry for Education and Research.

\appendix
\section{N=2 Supersymmetry Transformations} \label{sec:N2Susy}
In \cite{Sohnius}, the supersymmetry transformations of the fermions in the hypermultiplet are 
\begin{equation} \label{eq:ftrans}
\delta \Psi = -(i\gamma^\mu\partial_\mu +M+\gamma_5 N) \zeta^i B_i.
\end{equation}
and for the scalars they are 
  \begin{equation} \label{eq:transformation}
\delta B_i = 2\bar{\zeta_i} \Psi.
\end{equation} 
Here there is a symplectic Majorana condition on the SUSY parameters, defined by the condition
\begin{equation} \label{eq:symplectic}
\zeta^i = \epsilon^{ij} \gamma_5 C \bar{\zeta_j}^T,
\end{equation}
where C is given by 
\begin{equation}
C=\left(\begin{array}{cc}
-\varepsilon_{\alpha \beta} &  0 \\
0  & -\varepsilon^{\dot{\alpha} \dot{\beta}}
\end{array}\right),
\end{equation}
so in terms of Weyl spinors, the SUSY parameters are of the form
\begin{equation} 
\label{eq:WeylCon}
 \bar{\zeta}_1 =  \left(
  \begin{array}{cc}
 \chi_{\alpha} &  i\bar{\xi}^{\dot{\alpha}} 
  \end{array} \right),  \ \ \  
 \bar{\zeta}_2 =  \left(
  \begin{array}{cc}
  \xi_{\alpha} &  -i\bar{ \chi}^{\dot{\alpha}}
  \end{array} \right), \ \ \  
  \Psi =  \left(
  \begin{array}{c}
  \Psi_1 \\ \bar{\Psi}_2
  \end{array} \right).
 \end{equation}   
 Plugging this in the above relations yields the transformations quoted in section 2.
\comments{\section{Self-Energy Corrections} \label{sec:Divergence}
Here we will consider the self-energy corrections to the scalar propagator in the N=2 theory described in section 2. 
\comments{ We will consider the Wilson line defined in \eqref{eq:N2OWLDef}a Wilson line with the same scalar at either end
\begin{equation}
B_1 \textit{P}\exp (\int_C \, \,(iA_\mu dx^\mu + (v_5 M + v_6 N)|dx|) )  \bar{B_1}
\end{equation}}
Clearly the contribution from the gauge interaction with the scalar is the same as those found in \cite{Zarembo} up to gauge group indices, so we get
\begin{equation}
\delta_{ab}g^4C_2(r)\frac{\Gamma(2-\omega)\Gamma(\omega)\Gamma(\omega-1)}{(4\pi)^\omega \Gamma(2\omega)}4(2\omega-1)\frac{\delta_{ij}}{p^{6-2\omega}}.
\end{equation}
For the Fermion loop, we will also get the same result as in \cite{Zarembo}, which is 
\begin{equation}
-\delta_{ab}g^4C_2(r)\frac{\Gamma(2-\omega)\Gamma(\omega)\Gamma(\omega-1)}{(4\pi)^\omega \Gamma(2\omega)}8(2\omega-1)\frac{\delta_{ij}}{p^{6-2\omega}}.
\end{equation}
To see this, we can consider the N=4 theory with the gauge group SU(2)$_L$xSU(2)$_R$ broken to N=2 SYM with the gauge group SU(2)$_L$ as done in \cite{Pestun}.  In this case, we get an N=2 vector multiplet and an N=2 hypermultiplet.  If we choose scalars in the hypermultiplet, then their interactions with the fermions will have one vector and one hyper coupling.  More explicitly, the four chiral fermions of N=4 SYM are composed of one SU(2)$_L$ doublet in the vector multiplet $\lambda_i$ and one SU(2)$_R$ doublet in the hypermultiplet $\Psi_i$.  Since SU(2)$_R$ is broken, the doublet in the hypermultiplet becomes a single Dirac Fermion.  The Yukawa couplings of the N=4 theory with a hypermultiplet scalar $B_i$ are fixed by the global symmetry to have one of the fermions be $\lambda_i$ with the same SU(2)$_L$ index and to have the other a singlet of the full symmetry group.  This clearly singles out the coupling of the N=2 theory
\begin{equation}
-i\bar{\Psi}\lambda^i B_i+h.c.,
\end{equation}
and no others.  This means that all of the same couplings exist, and the contribution from the Fermion loop must be the same.  Similar argumentation applies to the scalar loops from four point couplings.  We therefore see that the self-energy corrections are the same for the N=2 theory as for the N=4 theory.}
\comments{
which is integrated over a semi-circular contour.  In the conventions used in this section, we will take the position space propagators for the scalars and gluons respectively to be 
\begin{eqnarray}
\Delta^a(x)=\frac{i}{4\pi^2 x^2}, \ \ \ \
\Delta^{\mu\nu}_A(y-z)=\frac{-ig^{\mu\nu}T_A}{4\pi^2 (y-z)^2}.
\end{eqnarray} 
There are three

In that paper, the scalar self energy contribution from the fermion loop was found to be $-2$ times the contribution from the gauge coupling, which is
\begin{equation}
\delta^{ab} g^4 C_2(r) \frac{\Gamma^2(\omega-1)}{2^4\pi^{2\omega}(2-\omega)(2\omega -3)} \frac{1}{[x^2]^{2\omega-3}}.
\end{equation}
\comments{
\begin{equxation}
\delta^{ab} g^4 N \frac{\Gamma(\omega-1)\Gamma(\omega)\Gamma(}{2^5\pi^{2\omega}(2-\omega)(2\omega -3)} \frac{1}{[x^2]^{2\omega-3}}
\end{equation}}
In our case there are only two fermions that contribute, instead of four.  Furthermore, since the coupling is from the term $iB^{\dagger i}\bar{\lambda}_i \Psi - i \bar{\Psi} \lambda^i B_i$ in the Lagrangian, the fermion from the vector multiplet is fixed.  This means there is a factor of $\frac{1}{4}$ relative to the result found in \cite{Zarembo}.  The result for the fermions is then 
\begin{equation}
-\delta^{ab} g^4 C_2(r) \frac{\Gamma^2(\omega-1)}{2^5\pi^{2\omega}(2-\omega)(2\omega -3)} \frac{1}{[x^2]^{2\omega-3}}.
\end{equation}
and the full expression for the self energy correction is 
\begin{equation}
\delta^{ab} g^4 C_2(r) \frac{\Gamma^2(\omega-1)}{2^5\pi^{2\omega}(2-\omega)(2\omega -3)} \frac{1}{[x^2]^{2\omega-3}}.
\end{equation}}
\comments{\\ \indent However, it can be proven that the diagram similar to that in figure \ref{fig:FirstOrder} diverges.  It is given by the expression 
\begin{equation} 
-\frac{1}{2}g^4 C_2(r)( \overleftarrow{\partial}_\nu^1 - \overrightarrow{\partial}_\nu^2 )\int dz_\mu \, d^4 y \,  \Delta(x_1-y) g_{ym}\Delta^{\mu\nu}(y-z)\Delta(x_2-y).
\end{equation}
We will not prove that this diverges here, but it is shown in \cite{Zarembo} that when $\{y,z\}\rightarrow x_1$ this type of integral diverges.  This can also be shown by a simple analysis near the point $x_1$.  This implies that at first order the operator diverges, which means it is not supersymmetric.  }
 \comments{ Since the only difference is that the scalar is in the fundamental representation instead of the adjoint representation, so we will just have to compute the difference in the prefactor.  Our self energy diagrams come from following terms in the Lagrangian:
\begin{equation} \label{eq:scalLag}
\frac{1}{2}A_\mu A_i^\dagger \partial^\mu A^i + iA^{\dagger i}\bar{\lambda}_i\Psi + h.c.
\end{equation}
So the gauge field structure of both of these diagrams is $\text{Tr}(T^a(r)T^a(r))=C_2(r)\cdot \bold{1}$, where the T's are in the fundamental representation.  The gauge structure in the N=4 theory gives an $f^{acd}f^{bcd}=C_2(G)\delta^{ab}$, so to switch from the results in \cite{Zarembo} to those required here we simply replace $C_2(G)=\frac{1}{2}N$ by $C_2(r)$ which is equal to $\frac{N^2-1}{2N}$ in the case of U(N). We thus get the following as the self-energy of the scalar
\begin{equation} \label{eq:scalarSelfEnergy}
-2\delta^{ab} g^4 C_2(r) \frac{\Gamma^2(\omega-1)}{2^5\pi^{2\omega}(2-\omega)(2\omega -3)} \frac{1}{[x^2]^{2\omega-3}}
\end{equation}
\indent The expression for the diagram in figure \ref{fig:orderOne} comes from the gauge coupling of the scalar to the gauge field, the first term in \eqref{eq:scalLag}.  The value is given by  
\begin{equation} \label{eq:pertExp1}
-\frac{1}{2}g^4 C_2(r)\int dz_\mu \, d^4 y \,  \Delta(x_1-y) g_{ym}( \overleftarrow{\partial}_\nu^1 - \overrightarrow{\partial}_\nu^2 )\Delta^{\mu\nu}(y-z)\Delta(x_2-y)
\end{equation}
it turns out that the divergences in \eqref{eq:scalarSelfEnergy} and \eqref{eq:pertExp1} cancel, which comes directly from arguments in \cite{Zarembo}, since they prove that two integrals of the same form cancel. In that case, they integrate over the positions of $x_1$ and $x_2$, but the result of this integral is the same for both cases. This means that to first order this diagram is finite.  We do not necessarily expect that these divergences will cancel at higher order, because the scalars did not come into play here, so we will not go into more detail in this calculation.  }
\section{Conventions} \label{sec:conventions}
We take the defect to be localized in the three direction, $x_3$.  In \cite{Defect} they use a Majorana basis defined by the following conventions for 4D $\gamma$ matrices and 3D ``$\rho$" matrices
\begin{eqnarray}
\rho^0=-\sigma^2, \ \ \ \ \rho^1=i\sigma^1, \ \ \ \ \rho^3=i\sigma^3 \ \ \ \ \\
\gamma^0 = \rho^0 \otimes \sigma^3, \ \ \ \ \gamma^1 = \rho^1 \otimes \sigma^3, \ \ \ \ \gamma^2 = \rho^2 \otimes \sigma^3, \ \ \ \ \gamma^3 = \text{I} \otimes i\sigma^1 \ \ \ \ \\
\gamma_5 = -i\gamma^0\gamma^1\gamma^2\gamma^3 = I\otimes \sigma^2. \ \ \ \ 
\end{eqnarray}  
In this basis, the second part of the direct product in the $\gamma$ matrices is used to reduce to 3D spinors, so that the top component and bottom component transform separately under $\gamma^k$ (k$\in\{0,1,2\}$) and are mixed under $\gamma^3$. Indeed, we see that under this decomposition we get
\begin{equation*}
\gamma_k  \left(
  \begin{array}{c}
  \lambda_1 \\
  \lambda_2
  \end{array} \right)
  = \left(
  \begin{array}{c}
  \rho^k \lambda_1 \\
  -\rho^k \lambda_2
  \end{array} \right)  , \ \ \ \ 
 \gamma_3  \left(
  \begin{array}{c}
  \lambda_1 \\
  \lambda_2
  \end{array} \right)
  = i\left(
  \begin{array}{c}
  \lambda_2 \\
   \lambda_1
  \end{array} \right).
\end{equation*}
This decomposition of $\lambda$ and $\chi_A$ can be used to decompose the bulk fields into 3D multiplets which transform independently under 3D supersymmetry, taking the bottom component of the supersymmetry transformation to be zero.  Under this reduction, the bulk vector multiplet splits into a three-dimensional vector multiplet consisting of $\{A_k,\lambda^1_{im},X^V_A\}$ and a three-dimensional hypermultiplet consisting of  $\{A_3,\lambda^2_{im},X^H_A\}$.  Here $\lambda_{im}$ is defined by the equation
\begin{equation} \label{eq:lamb}
 \lambda_{im}  =  \lambda \delta_{im} - i\sigma^A_{im}\chi_A, 
\end{equation}
where  $A,B,C\in \{1,2,3\}$, i and j are SU(2)$_V$ indices, and m and n are SU(2)$_H$ indices. 
\\The Lagrangian of the theory is composed of the usual N=4 Lagrangian coupled to the following defect Lagragian
\begin{eqnarray}
\frac{1}{g^2}\int d^3x ((D_kq^{m})^\dagger D^k q^m  - i\bar{\Psi}^i\rho^k D_k \Psi^i +\bar{f}^mf^m +i\bar{q}^m(\bar{\lambda}_1^a)_{mi}T^a\Psi^i -i\bar{\Psi}^i(\lambda_1^a)_{im}T^aq^m  \\ + \sigma_{mn}^I (\bar{q}^m(F_V^{Ia}-D_3 X_H^{Ia})T^aq^n+\bar{q}^mX_V^{Aa}T^af^n+\bar{f}^mX_V^{Aa}T^aq^n)). \nonumber
\end{eqnarray}
Solving for the auxiliary fields in this Lagrangian makes the full R symmetry group manifest, and is done in \cite{Defect}, but will not be necessary for us.
\section{DCFT Supersymmetry Transformations} \label{sec:DCFT}
In order to derive the supersymmetry transformations of the theory, we will use Majorana spinors.  The supersymmetry transformations for the N=4 theory in the bulk are \cite{Sohnius} (where some factors of i have been changed due to different conventions)
\begin{eqnarray} \label{eq:transformations}
\delta A_\mu = i\bar{\eta}_p \gamma_\mu \lambda^p \\
\delta X^V_{pq} = \bar{\eta_p} \lambda_q -\bar{\eta_q} \lambda_p + \epsilon_{pqrs} \bar{\eta}^r \lambda ^s \\
\delta X^H_{pq} = i(\bar{\eta_p} \gamma_5 \lambda_q -\bar{\eta_q}\gamma_5 \lambda_p - \epsilon_{pqrs} \bar{\eta}^r \gamma_5\lambda ^s),
\end{eqnarray}
where here p and q take values from 1 to 4, the X's are defined by the following equations
\begin{eqnarray}
X^V_{AB}=-\epsilon_{ABC}X^V_C; \ \ \ \ X^H_{AB}=-\epsilon_{ABC}X^H_C; \\
X^V_{4A}=-X^V_{A4}=X^V_A; \ \ \ \ \ X^H_{4A}=-X^H_{A4}=X^H_A;
\end{eqnarray}
and $\lambda_p$ is defined such that the gaugino $\lambda$ is the fourth component and the chiral spinors $\chi_A$ are the first three components.  According to the prescription for projecting spinors described above, $X_V$ and $X_H$ transform into the top and bottom components of the spinor respectively, as expected.  Since $A_3$ gives the transformation  
\begin{equation}
\delta A_3 = i\bar{\eta}_p \gamma_3 \lambda^p \rightarrow -\bar{\eta}^{top}_p \lambda_{bot}^p 
\end{equation}
we see that in order to cancel this supersymmetry transformation, we must choose a scalar also in the hypermultiplet, which for definiteness will be $X_3^H$:
\begin{equation} \label{eq:Btrans}
\delta X^H_3 = i(-\bar{\eta_1} \gamma_5\lambda_2 +\bar{\eta_2} \gamma_5\lambda_1 + \bar{\eta_3}\gamma_5 \lambda_4 -\bar{\eta_4}\gamma_5 \lambda_3 ).
\end{equation}
 The linear combination $\bar{\eta}_1 S_1+i\bar{\eta}_2 S_2+\bar{\eta}_3 S_3+i\bar{\eta}_4 S_4$ then annihilates the Wilson line if the $\eta$'s satisfy the conditions 
\begin{eqnarray} \label{eq:trans}
\bar{\eta}_1 \gamma_5 = \bar{\eta}_2 \gamma_3 \\
\bar{\eta}_3 \gamma_5 = -\bar{\eta}_4 \gamma_3, 
\end{eqnarray}
where there is an extra minus sign from the -$\varepsilon_{ABC}$ in the definition of the scalar.  This means that half of the generators annihilate the line, as expected.  We note that those relations imply that 
\begin{equation}
\bar{\eta}_1 \gamma_3 = \bar{\eta}_1 \gamma_5 \gamma_5 \gamma_3 = \bar{\eta}_2 \gamma_3 \gamma_5 \gamma_3 = \bar{\eta}_2 \gamma_5,
\end{equation}
which is consistent because when $i\bar{\eta}_2 S_2$ acts on $A_\mu$, the $i^2$ provides an extra minus sign.  
\\ \indent   To switch to the SU(2)$_V$xSU(2)$_H$ notation, we can define the SUSY generators as follows
\begin{equation} \label{eq:theta}
 S_{im}  =  S_4 \delta_{im} - i\sigma^A_{im}S_A, 
\end{equation}
similar to the definition of $\lambda_{ij}$ given in \eqref{eq:lamb}.  From the transformation
\begin{eqnarray}
S_4 X^H_C = -i\bar{\eta}_4 \gamma_5\chi_C 
\end{eqnarray}
 we can see that the linear combination $\bar{S}_4=\frac{1}{2}(\bar{S}_{11}+\bar{S}_{22})$ corresponds to the N=1 SUSY generator used in superspace in \cite{Defect}, and this is also true for the transformation of the vector multiplet $A_\mu$ (since the gaugino is the fourth component of $\lambda_i$).  
\\ \indent The transformation of the hypermultiplet scalars is 
\begin{eqnarray}
\delta q_m =2\bar{\zeta}_{im} \Psi_i \\
\delta \bar{q}_m =2\bar{\Psi}_i \zeta_{im}.
\end{eqnarray} 
\indent This means that if we choose, for example, the scalars $q_1$ and $\bar{q_1}$ at the ends of the line, we must demand
\begin{equation} \label{eq:cons2}
\bar{\zeta}_{i1}=\zeta_{i1}=0,
\end{equation}
but this is inconsistent, because in this basis the Majorana condition means that the $\eta$'s defined above are real, so \eqref{eq:theta} implies that $S^*_{im}=\varepsilon_{ij}\varepsilon_{mn}S_{jn}$.  Therefore a consistent choice would be to choose $q_1$ and $\bar{q_2}$ at the ends of the line.
\\ \indent To solve for the spinors which annihilate both the line and the scalars, we first restrict \eqref{eq:trans} to the 3d boundary, yielding
\begin{eqnarray} \label{eq:rel}
\zeta_1  = -\zeta_2, \ \ 
\zeta_3 = \zeta_4, 
\end{eqnarray}
where the index here has an implicit conversion as in \eqref{eq:theta}.
\\ \indent In fact, the condition \eqref{eq:rel} implies that the generator used above $\eta_1 S_1+i\eta_2 S_2$, subject to \eqref{eq:rel}, can be rewritten using \eqref{eq:theta} as $i\zeta_{12}S_{12}$ with $\zeta$ Majorana.  Similarly, the generator $\eta_3 S_3+i\eta_4 S_4$ can be written as $i\zeta_{22}S_{22}$. This means we should choose $\bar{q}_1$ at one end of the line, and since $\bar{S}_{im}$ is related to $S_{im}$ through $S^*_{im}=\varepsilon_{ij}\varepsilon_{mn}S_{jn}$, we must then choose $q_2$ at the other end of the line for the full variation of the operator to vanish.

\bibliographystyle{ieeetr}
\bibliography{SusyTransformations}
\end{document}